\ifx\empty\selectedoptions
  \def\selectedoptions{final}
\fi

\documentclass[]{aipproc}

\def\ao{Appl.\  Opt.\ }
\def\aap{Astron.\  Astroph.\ }

\def\apj{Astrophys.\ J.\ }
\def\aj{Astron.\ J.\ }
\def\apjl{Astrophys.\ J.\ Lett.\ }

\def\pasp{Proc.\ Ast.\ Soc.\ Pac.\ }

\newfont{\rmsmall}{cmr10 scaled 900}
\newcommand{\micron}{\mu{\mathrm{m}}}

\newcommand{\ci}{C~{\rmsmall I}}
\newcommand{\gs}{{\mathrel{\raise0.35ex\hbox{$\scriptstyle
>$}\kern-0.6em
\lower0.40ex\hbox}{{$\scriptstyle \sim$}}}}
\newcommand{\ls}{{\mathrel{\raise0.35ex\hbox{$\scriptstyle
<$}\kern-0.6em
\lower0.40ex\hbox}{{$\scriptstyle \sim$}}}}

\layoutstyle{8x11single}

\begin{document}

\title 
      [South Pole Observations]
      {Millimeter and Submillimeter Observations from the~South~Pole}


\author{Antony A. Stark}{
  address={Smithsonian Astrophysical Observatory, 60 Garden St.,
Cambridge, MA 02138 USA},
  email={aas@cfa.harvard.edu},
  thanks={also associated with the Center for Astrophysical Research
in Antarctica}
}

\copyrightholder{American Institute of Physics}
\copyrightyear  {2001}

\begin{abstract}
During the past decade, a year-round observatory has been established
at the geographic South Pole by the Center for Astrophysical Research in
Antarctica  (CARA).  CARA has fielded several millimeter- and
submillimeter-wave instruments:  AST/RO (the Antarctic Submillimeter
Telescope and Remote Observatory, a 1.7-m telescope outfitted with
a variety of receivers at frequencies from 230 GHz to 810 GHz,
including PoleSTAR, a heterodyne spectrometer array), Python (a
degree-scale CMB telescope), 
Viper (a 2-m telescope which has been outfitted with
SPARO, a submillimeter-wave bolometric array polarimeter, ACBAR, 
a multi-channel CMB instrument, and Dos Equis, a HEMT polarimeter),
and DASI (the Degree-Angular Scale Interferometer).  These
instruments have obtained significant results in studies of the
interstellar medium and observational cosmology,
including detections of the $1^{\circ}$ acoustic peak in the CMB and the
Sunyaev-Zel'dovich effect.  The South Pole environment is unique among
observatory sites for unusually low wind speeds, low absolute humidity,
and the consistent clarity of the submillimeter sky.  The atmosphere is
dessicated by cold: at the South Pole's average annual temperature of
-49 C, the partial pressure of saturated water vapor is only 1.2\% of
what it is at 0 C.  The low water vapor levels result in exceptionally
low values of sky noise.  This is crucial for large-scale observations
of faint cosmological sources---for such observations the South Pole is
unsurpassed. 
\end{abstract}

\date{\today}

\maketitle

\section{Introduction}

Development of the geographic South Pole as an astronomical observatory
has largely been driven by the desire to measure
anisotropy in the Cosmic Microwave Background (CMB) radiation.
Observers have actively pursued the measurement of CMB
anisotropy because of the theoretical expectation
that those
measurements would significantly
advance our understanding of the Universe \citep{peebles71}.
Deep millimeter- and submillimeter-wave
observations, of which CMB anisotropy measurements
are an extreme example,
are often made impossible by opacity 
and noise due to water vapor in the 
Earth's atmosphere \citep{radford86}.
For large angular scale measurements, this
problem has been successfully finessed by
orbital \citep{smoot92} and airborne \citep{debernardis00,hanany00} instruments,  but
observations of anisotropy on scales  
less than a few arcminutes
require large telescope apertures, which
cannot be deployed above the Earth's surface using
current technology.
This hard truth has led observers to seek 
new observatory sites having the smallest possible
atmospheric water vapor while still being practical locations
for the installation and operation of large telescopes.
Among the most promising of the new observatories is
Amundsen-Scott South Pole Station, the permanent 
U.S. National Science Foundation base on the
Antarctic Plateau.

South Pole Station was originally built for the International
Geophysical Year in 1957 and has been in continuous use
ever since.  Station infrastructure is currently being rebuilt for
the second time; the new station is scheduled for completion
in 2005.  The station supports 50 scientists and
staff throughout the Austral winter, with an increase to
over 200 during the summer months.
From the earliest days, scientific investigations at South Pole Station included
observations of  weather \citep{schwerdtfeger,warren96} and 
atmospheric phenomena such as
aurora \citep{landolt58} and ice halos \citep{ohtake78}.  The South Pole
has evolved into
an important site for seismology \citep{roult94}, solar astronomy \citep{harvey89,libbrecht91},
atmospheric studies \citep{fan99}, and cosmic ray physics \citep{smith89}.   

Because it is exceptionally cold, 
the climate at the Pole implies exceptionally dry observing conditions.
As air becomes 
cold, the amount of water vapor it can
hold declines dramatically.
Air at 0 C can hold 83 times more water vapor than saturated air at 
the South Pole's average annual temperature
of $-49$ C \citep{goff46}; 
together with the relatively high altitude 
of the Pole (2850 m), this means 
the water vapor content of the atmosphere above the South
Pole is two or three orders of magnitude smaller than 
it is at most places on the Earth's surface.
This has long been known
\citep{smythe77}, but many years of hard work were needed to
realize the potential in the form of
new astronomical knowledge.

The EMILIE experiment by the French group \citet{pajot89} made the 
first astronomical observations
of  submillimeter-waves
from the South Pole during the Austral summer of 1984-1985.
EMILIE was a ground-based single-pixel bolometer dewar 
operating at $900 \mu \mathrm{m}$ and
fed by a 45 cm off-axis mirror.
It had successfully measured the diffuse galactic emission 
while operating on Mauna Kea in Hawaii in 1982, but
the accuracy of the result had been limited by sky noise \citep{pajot86}.   
Martin A. Pomerantz, a cosmic ray researcher at Bartol Research Institute, and
John T. Lynch, the 
NSF Program Director for Antarctic Aeronomy and Astrophysics \citep{lynch98}, encouraged
the EMILIE group to relocate their experiment to the South Pole.
There they found better observing conditions and were able to make
improved measurements of galactic emission \citep{pajot89a}.

Pomerantz also enabled Mark Dragovan, then a
researcher at Bell Laboratories, to attempt CMB anisotropy
measurements from the Pole.  
Dragovan, Stark, and Pernic \citep{dragovan90} 
built a lightweight 1.2 m offset telescope and were
able to get it working at the Pole with a single-pixel
helium-4 bolometer during several weeks in January 1987.
The results were sufficiently encouraging
that several CMB groups \citep{dragovan90a,gaier89,meinhold89,peterson89} 
participated in the ``Cucumber'' campaign in
the Austral summer of 1988-1989, where three Jamesway tents and a 
generator 
dedicated to CMB anisotropy measurements
were set up at a temporary site 2 km from South Pole Station.
These were summer-only campaigns, where instruments were shipped in,
assembled, tested, used, disassembled, and shipped out in a single 
three-month-long summer season.  Considerable time and effort were expended in 
establishing and then demolishing observatory facilities, with little
return in observing time.  What little observing time was available occurred
during the warmest and wettest days of mid-summer.

Permanent, year-round facilities were needed.
The Antarctic Submillimeter Telescope and Remote Observatory 
(AST/RO) \citep{stark97a,stark01}
is a 1.7 m diameter offset gregorian
telescope mounted on a dedicated, permanent observatory building.
It was the first radio telescope to operate year-round at South Pole. 
AST/RO was started in 1989 as an independent project, but
in 1991 it became part of a newly-founded National Science Foundation
Science and Technology Center, the Center for Astrophysical Research
in Antarctica (CARA)
\citep[see also  \url{http://astro.uchicago.edu/cara}]{landsberg98}.
CARA has fielded several telescopes in addition to
AST/RO:
White Dish \citep{tucker93}, 
Python \citep{dragovan94,alvarez95,ruhl95,platt97,coble99} and 
Viper \citep{peterson00} (Cosmic Microwave Background experiments),
DASI \citep{leitch01a} (the Degree-Angular Scale Interferometer), and 
SPIREX \citep{nguyen96} (the South Pole Infrared Explorer, 
a 60-cm telescope, now decommissioned).
These facilities are housed in the ``Dark Sector'', a grouping of
buildings which includes
the AST/RO building, the Martin A. Pomerantz Observatory building (MAPO) and
a new ``Dark Sector Laboratory'',
all located 1 km away from the main base across the aircraft runway in a
radio quiet zone.

\begin{figure}[tb!]
\includegraphics[width=\textwidth]{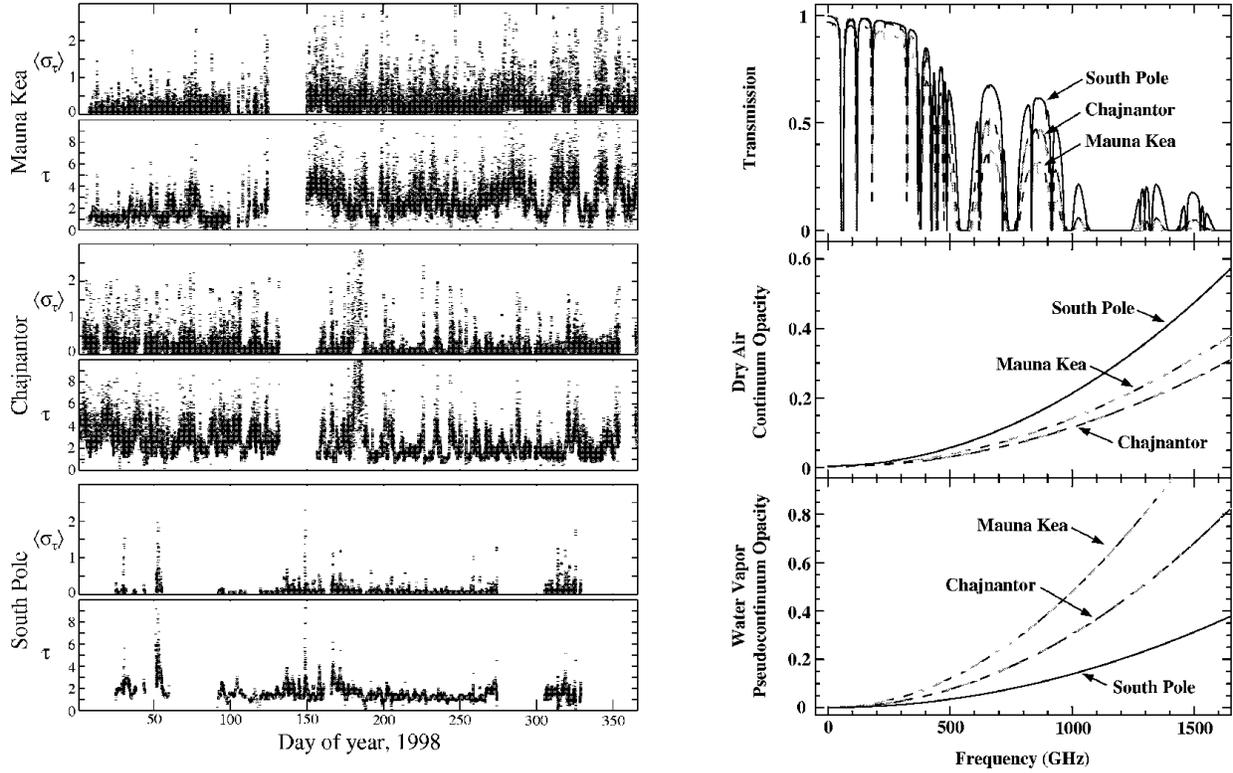}
\caption{
Opacity and sky noise at the South Pole.
(Left) Measurements at $ 350 \, \mu\mathrm{m} $ from three
sites.
These plots show data from identical NRAO-CMU $ 350 \, \mu\mathrm{m} $  broadband
tippers
located at Mauna Kea, the ALMA site at Chajnantor in Chile, and South Pole
during 1998.  The upper
plot of each pair shows $\langle \sigma_\tau \rangle$, the rms
deviation
in the opacity $\tau$ during a one-hour period---a measure
of sky noise on large scales; the lower plot of each
pair shows $\tau$, the broadband $350 \, \mu\mathrm{m}$  opacity.
The first 100 days of 1998 on Mauna Kea were exceptionally good for
that site, due to a strong El Ni\~{n}o that year.
During the best weather at the Pole,
$\langle \sigma_\tau \rangle$ was dominated by detector noise rather
than
sky noise.
(Right) Calculated atmospheric transmittance at the three sites. 
The upper plot is atmospheric transmittance at zenith
calculated by J. R. Pardo using the ATM model \cite{pardo01b}.
The model uses PWV values of 0.2 mm for South Pole, 0.6 mm
for Chajnantor and 0.9 mm for Mauna Kea, corresponding to
the $25^{\mathrm{th}}$ percentile winter values at each site.
Note that at low frequencies, the Chajnantor curve converges with the
South Pole curve, an indication that 225 GHz opacity is not a simple
predictor of submillimeter wave opacity.
The middle and lower plots show calculated values of dry air
continuum opacity and water vapor pseudocontinuum opacity for
the three sites.  Note that unlike the other sites, the opacity
at South Pole is
dominated by dry air rather than water vapor.
\label{fig:combinedsite}
}
\end{figure}

\section{Site Testing}

The South Pole is an excellent millimeter- and submillimeter-wave site \citep{lane98}.
It is unique among observatory sites for unusually low
wind speeds, the complete absence of rain, and the consistent clarity of the
submillimeter sky.  
\citet{schwerdtfeger}
has comprehensively reviewed the climate of the Antarctic Plateau
and the records of the
South Pole meteorology office.
\citet{chamberlin01}
has analyzed weather data to determine the precipitable
water vapor (PWV) and finds median wintertime 
PWV values of 0.3 mm over a 37-year period, with little annual
variation.
{\em PWV values
at South Pole are small, stable, and well-understood.}

Sub\-millimeter-wave atmospheric
opacity at South Pole has been measured using skydip techniques.
\citet{chamberlin97} made
over 1100 skydip observations at 492~GHz (609 $\micron$ ) with AST/RO
during the 1995 observing season.
Even though this frequency is near a strong oxygen line, the opacity
was below 0.70 half of the time during the Austral winter and reached
values as
low as 0.34, better than ever measured at any other ground-based site. 
From early 1998, the 350 $\micron$ band has been continuously
monitored
at Mauna Kea, Chajnantor, and South Pole by identical tipper
instruments
developed by S. Radford of NRAO and J. Peterson of Carnegie-Mellon U.
and
CARA.  Results from Mauna Kea and Chajnantor are compared with South Pole in
the left panel of Figure~\ref{fig:combinedsite}.
{\em The 350 $\micron$ opacity at the South Pole is consistently better
than at Mauna Kea or Chajnantor.}
The South Pole 25\% winter PWV levels
have been used to compute values of atmospheric
transmittance
as a function of wavelength and are plotted in the right panel of
Figure~\ref{fig:combinedsite}. For comparison, the transmittance for
25\% winter conditions at Chajnantor and Mauna Kea is also shown.

{\em Sky noise} is fluctuations
in total power or phase of a detector caused by
variations in atmospheric emissivity and path length on timescales of
order one second.
Sky noise causes systematic errors in the measurement of astronomical
sources.  
\citet{lay99} 
show analytically how sky noise causes observational
techniques
to fail: fluctuations in a component of the data due to sky noise
integrates
down more slowly than $t^{-1/2}$
and will come to dominate the error during long
observations. 
Sky noise at South Pole is considerably smaller than at other sites,
even comparing conditions of
the same opacity.  
The PWV at South Pole is
often so low that the opacity is dominated by the {\em dry air}
component \citep[cf. Figure \ref{fig:combinedsite}]{chamberlin95,chamberlin01};
the dry air emissivity
and phase error do not vary as strongly or rapidly as
the emissivity and phase error due to water vapor.
\citet{lay99} 
have compared
the Python experiment at South Pole
with the Site Testing Interferometer
at Chajnantor \citep{radford96,holdaway95}
and find that the amplitude of the sky noise at
South Pole is 10 to 50 times less than that at Chajnantor.
{\em The strength of South Pole as a millimeter and submillimeter site lies in
the low sky noise levels routinely obtainable for sources
around the South Celestial Pole.}

\section{Logistics}

South Pole Station provides logistical support for observatory experiments: room and
board for on-site
scientific staff, electrical power, network and telephone connections,
heavy equipment support, and cargo and personnel transport.
The station powerplant provides about 100 kW of power to
CARA projects out of a total generating capacity of about 600
kW.
Heavy equipment at South Pole Station
includes cranes, forklifts, and bulldozers;
these can be requisitioned for scientific use as needed.
The station is supplied by over 200
flights each year of LC130 ski-equipped cargo
aircraft.  Annual cargo capacity is about 3500 tons.
Aircraft flights are scheduled only during the period
from late October to early February so that the station
is inaccessible for as
long as nine months of the year.
All engineering operations for equipment installation and
maintenance are tied to the annual cycle of physical access to the
instruments.
For quick repairs and upgrades during the Austral
summer season, it is possible to send
equipment between South Pole and anywhere serviced by
commercial express delivery in about
five days.
During the winter, however, no transport is possible and projects
must be designed and managed accordingly.

In summer, there are about 20 CARA people at the
Pole at any given time.
Each person stays at Pole for a few weeks or months in order to
carry out their
planned tasks as well as circumstances allow, then they depart
to be replaced by another CARA person.
Each year there are four or five CARA winter-over scientists,
who remain at the observatory for a year.

The receivers used on AST/RO and Viper require about 20 liters of liquid
helium per day.  
Helium also escapes from the station in
one or two 
weather balloons launched each day.
The National Science Foundation and its support contractors
must supply helium to South Pole, and
the most efficient way to transport and supply helium is in liquid
form.
Before the winter-over period, one or more large (4000 to 12000 liter)
storage
dewars are brought to
South Pole for winter use; some years this supply lasts the
entire winter, but in 1996,
2000, and 2001 it did not.  
The supply of liquid helium has been a chronic problem
for South Pole astronomy, but improved facilities in
the new
station should eliminate single points of failure and provide
a more certain supply.

Internet and telephone service to South Pole is
provided by a combination of two low-bandwidth satellites, LES-9 and
GOES-3,
and the high-bandwidth (3 Mbps) NASA
Tracking and Data Relay Satellite System TDRS-F1.
These satellites are geosynchronous but not geostationary, since
their orbits are inclined.
Geostationary satellites are always below the horizon and cannot be
used.
Internet service is intermittent through each 24-hour period because
each satellite is visible only during the southernmost part
of its orbit; the combination of the three satellites provides an
Internet
connection for approximately 12 hours within the period 1--16 hr
Greenwich LST.
The TDRS link helps provide a
store-and-forward automatic transfer service for large computer files.
The total data communications capability is about 5 Gbytes per day.
Additional voice communications are provided by a fourth satellite,
ATS-3,
and high frequency radio.

\begin{figure}
\includegraphics[height=.35\textheight]{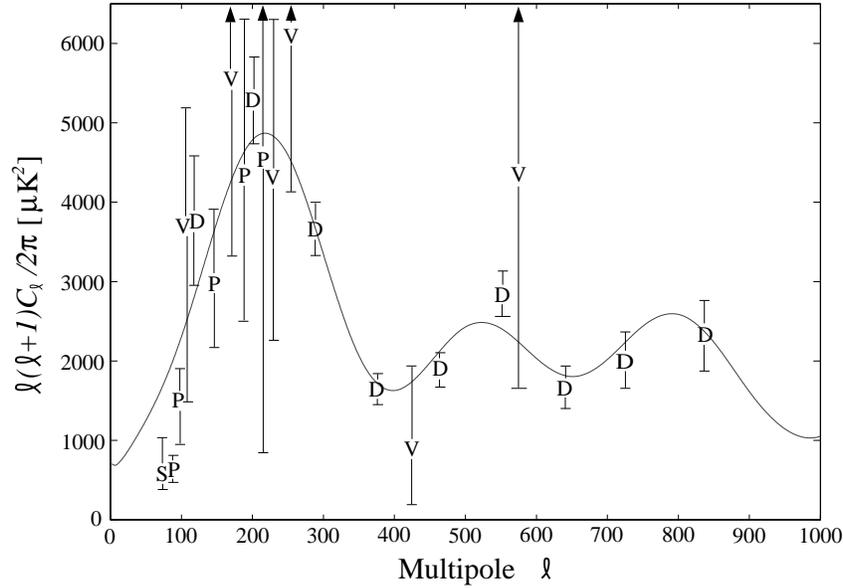}
\caption{
Summary of Cosmic Microwave Background Anisotropy measurements from the
South Pole.  The data shown are: S = U. of California, Santa
Barbara \citep{ganga97}; P = Python 1991-1996 \citep{coble99};
V = Viper 1998 \citep{peterson00}; D = DASI 2000 \citep{halverson01}.
The curved line is not a fit to the data, but instead shows the 
expected values for the cosmological
``concordance model'': $\Omega_{\mathrm{b}} = 0.05$,
$\Omega_{\mathrm{cdm}} = 0.35$,
$\Omega_{\Lambda} = 0.60$,
$\tau_{\mathrm{c}} = 0$,
$n_{\mathrm{s}} = 1.00$,
$H_{0} = 65 {\, \mathrm{km \, s^{-1} \, kpc^{-1}}}$
\citep{pryke01}.
The data are in remarkably good agreement with the concordance model,
and strongly reject many other cosmologies.
}
\label{fig:spcmbr}
\end{figure}

\begin{figure}
\includegraphics[height=.41\textheight]{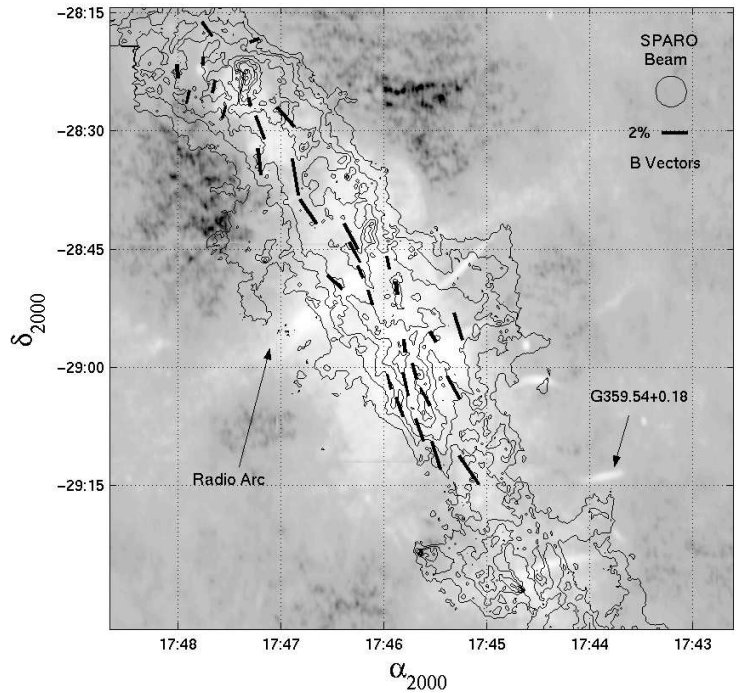}
\caption{
SPARO polarimetry results from \citet{novak01}
for the Galactic Center Region (vectors) superposed on
radio continuum map by \citet{larosa00} (gray scale).  The length of
each vector gives the measured degree of polarization (see key at upper
right), and the orientation of each vector shows the direction of the
inferred magnetic field, which is orthogonal to the measured direction of
polarization.  Black contours show the 800 $\micron$ dust emission measured using
the JCMT \citep{pierceprice00}.  The radio continuum map shows
several of the non-thermal filaments that trace magnetic field lines in
hot ionized regions.
}
\label{fig:sparo}
\source{G. Novak}
\end{figure}

\begin{figure}
\includegraphics[height=.31\textheight]{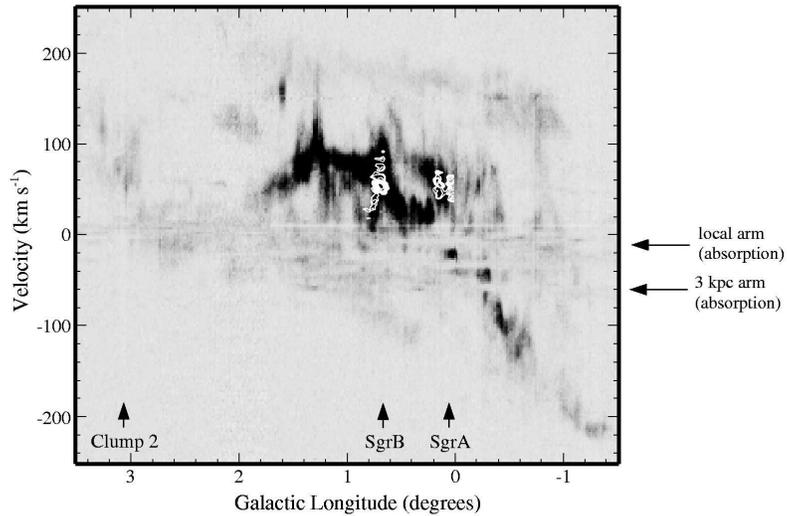}
\caption{
AST/RO observations of the Galactic Center Region
from \citet{kim00}.
The CO $J = 4 \rightarrow 3$ (greyscale) and $J = 7 \rightarrow 6$ 
(white contour) lines observed in an
$l - v$ strip, sampled every $1'$, at $b = 0^{\circ}$.
These data have been used in conjunction with
CO  and $^{13}$CO $J = 1 \rightarrow 0$ data to determine the
the density and temperature in the features shown here
\citep{kim00}.  The
300 parsec ring is the rough parallelogram-shape between $l = -1^{\circ}$ and
$l = 1.8^{\circ}$, and velocities from $-200$ to $+200 \, \mathrm{km \, s^{-1}}$.
The ring is just below the critical density at which
instabilities will cause it to coagulate into a single cloud
like Sgr B and Sgr A.
}
\label{fig:astrodata}
\source{S. Kim}
\end{figure}

\section{Current Instruments and Illustrative Results}

The Degree Angular Scale Interferometer (DASI) 
\citep{leitch01a}
is a compact centimeter-wave
interferometer designed to image the CMB primary anisotropy and measure its
angular power spectrum and polarization at angular scales ranging from two degrees to 
several arcminutes.  As an interferometer, DASI measures CMB power by simultaneous 
differencing on several scales, measuring the CMB power spectrum directly.
DASI was installed on a tower adjacent to the MAPO during the
1999-2000 Austral summer and has had two successful winter seasons so far.
The first season results are a stunning confirmation of
the ``concordance''
cosmological model, which has a flat geometry and significant contributions
to the total stress-energy from dark matter and dark energy
\citep{halverson01,pryke01}.

Viper is a 2.1 meter off-axis telescope designed to allow measurements of low
contrast astronomical sources.  It is 
mounted on another tower at the opposite end of the MAPO from DASI.
Viper is used with a variety of instruments:
Dos Equis, a CMB polarization receiver operating at 7 mm, SPARO,
a bolometric array polarimeter operating at 450 $\mu \rm m$, and ACBAR, a 
multi-wavelength bolometer array used to map the CMB.

ACBAR is a 16-element bolometer array operating at 300 mK.  It was designed
specifically for observations of CMB anisotropy and the Sunyaev-Zel'dovich effect
(SZE).  It was installed on the Viper telescope during the Austral
summer of 2000-2001 and was successfully operated throughout most of the Austral
winter of 2001.  ACBAR has made high-quality maps of SZE in several
nearby clusters of galaxies and has made significant measurements of
anisotropy on the scale of degrees to arcminutes.

The Submillimeter Polarimeter for Antarctic Remote Observing (SPARO)
is a 9-pixel polarimetric imager operating at 450 $\mu \rm m$.
It was operational on the Viper telescope during the early Austral winter of 2000.
\citet{novak01} mapped the polarization of an extended region of the sky ($ \sim 0.25$
square degrees) centered approximately on the Galactic Center.  Their
results imply that, within the Galactic Center molecular gas complex, the
toroidal component of the magnetic field is dominant. 
The data show
that all of the existing observations of large-scale magnetic fields in
the Galactic Center are basically consistent with the ``magnetic outflow''
model of \citet{uchida85}.  This magnetodynamic model was
developed in order to explain the Galactic Center Radio Lobe, a
limb-brightened radio structure that extends up to one degree above the
plane and may represent a gas outflow from the Galactic Center.

AST/RO is general-purpose 1.7 m diameter telescope \citep{stark97a,stark01} for astronomy
and aeronomy studies at wavelengths between
200 and 2000 $\mu \rm m$.
It is used primarily for spectroscopic studies of neutral atomic
carbon and carbon monoxide in the interstellar medium of the Milky Way
and the Magellanic Clouds
(see reference list at \url{http://cfa-www.harvard.edu/~adair/AST_RO}).
Five heterodyne
receivers and three acousto-optical spectrometers are
currently installed:
(1) a 230 GHz SIS receiver
\citep{kooi92},\
(2) a 450--495 GHz SIS quasi-optical receiver
\citep{engargiola94,zmuidzinas92},\
(3) a 450--495 GHz SIS waveguide
receiver \citep{walker92,kooi95}, which can be
used simultaneously with
(4) a 800-820 GHz fixed-tuned
SIS waveguide mixer receiver
\citep{honingh97}, and 
(5) the PoleSTAR array---four 800-820 GHz fixed-tuned
SIS waveguide mixer receivers
\citep[see \url{http://soral.as.arizona.edu/pole-star}]
{walker01,groppi00}.
Spectral lines observed with AST/RO include:
CO $J = 7 \rightarrow 6$,
CO $J = 4 \rightarrow 3$,
CO $J = 2 \rightarrow 1$,
HDO $J = 1_{0,1} \rightarrow 0_{0,0}$,
[\ci\,] $^3P_1 \rightarrow {}^3P_0$,
[\ci\,] $^3P_2 \rightarrow {}^3P_1$, and
[${}^{13}$\ci\,]  $^3P_2 \rightarrow {}^3P_1$.
There are four currently available
acousto-optical spectrometers (AOS) 
\citep{Schieder89}:
two low resolution spectrometers
with a bandwidth of 1 GHz,
an array AOS having four low
resolution spectrometer channels with a bandwidth of 1 GHz
for the PoleSTAR array,
and
one high-resolution AOS with 60 MHz bandwidth.
Two new instruments for AST/RO are under development:
TREND, a 1.5 THz heterodyne receiver \citep{gerecht99,yngvesson01}
and SPIFI, an imaging Fabry-Perot interferometer \citep{swain98}.
AST/RO observing time is open to all astronomers on a proposal basis.

\section{The Future 8-m Telescope}

An 8 m diameter off-axis telescope 
has been proposed for the South Pole
\citep{nrc01}.
It will be equipped with a large field of view \citep{stark00} that
will feed a
state-of-the-art 1027-element bolometer array receiver.
The initial science goal will be a large 
SZE survey covering 4000 square degrees at $1.3'$ resolution with
$10\,\mu$K sensitivity at a wavelength of 2~mm.  The SZE survey will
find all galaxy clusters
above a mass limit of
$3.5 \times 10^{14}$~M$_\odot$ regardless of redshift; should clusters
exist at redshifts much higher than currently predicted, they will be
found by the SZE survey, but missed in even the deepest X-ray
observations planned. It is expected that an unbiased sample
of approximately 20,000 clusters will be found, with over 1,000
at redshifts greater than one.  The sample will provide sufficient
statistics to use the density of clusters to determine the
equation of state of the dark energy component of the Universe.

\begin{figure}
\includegraphics[height=.25\textheight]{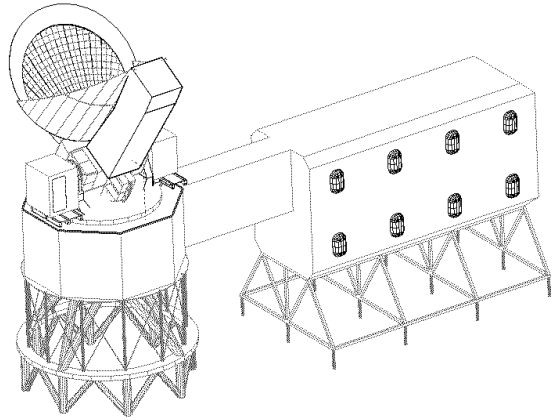}
\caption{ \label{fig:spt}
A sketch of the proposed 8-m South Pole Telescope and tower mounted
to the side of the new Dark Sector Laboratory. A guard
ring surrounding the primary and shielding around
the lower part of the secondary arm support region
minimize contamination from scattered radiation. The
guard ring allows the full aperture to be illuminated
without excessive noise from spillover. Like the
DASI and AST/RO telescopes, the environment
of all critical and moving
components will be kept warm, and therefore access will
not require the winterover staff to work in the extreme
cold conditions.}
\end{figure}

\begin{theacknowledgments}
We thank 
Jeff Peterson of CMU 
and 
Simon Radford of NRAO 
for the data
shown in the left panel of Figure \ref{fig:combinedsite}.
We thank Juan R. Pardo of Caltech for discussions on
atmospheric modeling and for carrying out
the calculations shown in the right panel of 
Figure \ref{fig:combinedsite}.
This work was supported in part 
by
the Center for Astrophysical Research in Antarctica and the NSF
under Cooperative Agreement OPP89-20223.
\end{theacknowledgments}

\hyphenation{Post-Script Sprin-ger}

\end{document}